\def\RELEASE{0}  %
\def\ANON{0}     %
\def\SQUEEZE{0}  %

\documentclass[sigconf,10pt]{acmart}
\PassOptionsToPackage{hyphens}{url}\usepackage{hyperref}

\setcopyright{none}
\renewcommand\footnotetextcopyrightpermission[1]{} %
\acmSubmissionID{0} %
\acmConference[HotOS'25]{HotOS}{May 2025}{Banff, Canada}

\usepackage[noend]{algpseudocode}
\usepackage{algorithmicx}
\usepackage{booktabs}
\usepackage{caption}
\usepackage{comment}
\usepackage[inline]{enumitem}
\usepackage{graphicx}
\usepackage{listings}
\usepackage{multirow}
\usepackage{xcolor}
\usepackage{xspace}
\usepackage{libertine}
\usepackage{microtype}
\usepackage{tikz}
\usepackage{gnuplot-lua-tikz}

\microtypecontext{spacing=nonfrench}

\hypersetup{
  pdflang={en},
  pdfdisplaydoctitle}

\definecolor[named]{OurPurple}{cmyk}{0.55,1,0,0.15}
\definecolor[named]{OurDarkBlue}{cmyk}{1,0.58,0,0.21}
\hypersetup{
  colorlinks,
  linkcolor=OurPurple,
  citecolor=OurPurple,
  urlcolor=OurDarkBlue,
  filecolor=OurDarkBlue}

\if \SQUEEZE 1
\setlist[itemize]{
  leftmargin=*,
  itemsep=2pt,
  topsep=2pt}

\fi
\usepackage{relsize}
\newcommand{\circled}[1]{{\small\protect\raisebox{0.5pt}{\textcircled{\raisebox{-.2pt}{\textls[-50]{\relsize{-1.5}\phantom{0}\makebox[0pt][c]{#1}\phantom{0}}}}}}\xspace}

\def\Snospace~{\S{}}

\if \RELEASE 1
  \def\NOTES{0}
\else
  \def\NOTES{1}
\fi
\if \NOTES 1
  \newcommand{\XXX}[1]{{\color{red}{XXX {#1}}}}
  \newcommand{\antoine}[1]{{\color{teal}{[\textbf{AK:} {#1}]}}}
  
  \newcommand{\vaas}[1]{{\color{orange}{[\textbf{VA:} {#1}]}}}
  \newcommand{\todo}[1]{{\color{blue}{TODO: {#1}}}}
\else
  \newcommand{\XXX}[1]{}
  \newcommand{\antoine}[1]{}
  \newcommand{\vaas}[1]{}
  \newcommand{\authorC}[1]{}
  \newcommand{\todo}[1]{}
\fi

\if \ANON 1
  \newcommand{\sys}{SystemX\xspace}
\else
  \newcommand{\sys}{LaissezCloud\xspace}
\fi

\begin{document}
\date{}
\title{It’s the People, Not the Placement: Rethinking Allocations in Post-Moore Clouds}

\if \ANON 1
  \author{Anonymous Submission \#XXX (\pageref{page:last pages})}
\else
  \author{
    Tejas S. Harith\\
    Max Planck Institute for Software Systems
    \and
    Antoine Kaufmann\\
    Max Planck Institute for Software Systems
    }
\fi

\begin{abstract}
The Cambrian explosion of new accelerators, driven by the slowdown of Moore’s Law, has created significant resource management challenges for modern IaaS clouds.
Unlike the homogeneous datacenters backing legacy clouds, emerging \emph{neoclouds} amass a diverse portfolio of heterogeneous hardware---NVIDIA GPUs, TPUs, Trainium chips, and FPGAs.
Neocloud operators and tenants must transition from managing a single large pool of computational resources to navigating a set of highly fragmented and constrained pools.
We argue that cloud resource management mechanisms and interfaces require a fundamental rethink to enable efficient and economical neoclouds.
Specifically we propose shifting from long-term static resource allocation with fixed-pricing to dynamic allocation with continuous, multilateral cost re-negotatiaton.
We demonstrate this approach is not only feasible for modern applications but also significantly improves resource efficiency and reduces costs.
Finally, we propose a new architecture for the interaction between operators, tenants, and applications in neoclouds.
\end{abstract}
 
\maketitle

\section{Introduction}
\label{sec:intro}

Traditional CPU-centric IaaS clouds are built on a backbone of homogeneous hardware.
When tenants request to spin up a new VM, operators can easily allocate resources from any machine in the pool.
If necessary, VMs can be seamlessly migrated between physical hosts without tenants noticing, enabling efficient resource utilization.
Transparent migration also simplifies operations by providing the flexibility needed for tasks such as maintenance, power, and cooling management.
This operational and resource efficiency ultimately results in lower prices for tenants.

In contrast, neoclouds, such as Coreweave, Lambda, and Crusoe~\cite{company:coreweave, company:lambdalabs, company:crusoe} are built on a backbone of heterogeneous platforms. 
These platforms feature diverse architectures and software interfaces, resulting in a highly fragmented ecosystem.
Tenants must specify concrete hardware platforms at a fixed price up-front when starting instances.
Once instances are started, transparent workload migration becomes either impossible or extremely costly for operators and tenants, as it is limited to small pools of identical hardware.

Without the ability to migrate workloads across platforms, neocloud operators are constrained by their initial allocations and must rely on conservative predicted usage patterns to size their hardware pools and allocate resources.
This approach often leads to suboptimal local maxima, where client requests are allocated in the order they arrive and remain static for the application’s lifetime.
Further complicating the issue, application performance varies significantly across different accelerators~\cite{bykov:dlchap, reagan:machsuite}, making it particularly challenging to manage resources in an ecosystem of rapidly evolving hardware and workloads~\cite{mirhoseini:devplace}.
Emerging AI neoclouds attempt to address this by working closely with tenants to design coarse-grained contracts that align applications with their most suitable hardware.
However, this approach results in poor resource efficiency, increased risk for operators, higher costs for tenants, and reduced convenience and flexibility.

In this paper, we argue that neoclouds require a fundamentally different approach to achieve efficient resource management, lower costs, and better flexibility for both operators and tenants. 
To this end, we propose a dynamic approach centered on renegotiating allocations and pricing online as client workloads are running. 
By building a framework for applications to evaluate their current runtime characteristics in the context of fluctuating cloud resource availability, we enable the co-design of clients' cost-performance objectives and cloud operators' management objectives.

Preliminary evaluations demonstrate that this approach can be implemented with manageable complexity for applications while significantly improving resource efficiency—both locally for individual tenants and globally across the cloud ecosystem. %
\section{Static Allocation and Pricing Is Inefficient}%
\label{sec:motivation}

Neoclouds suffer from poor resource and cost efficiency due to the sequential resolution of fixed allocations.
When a tenant arrives, it can only be allocated available instances, which it then holds until termination.
Moreover, accelerated applications often experience non-uniform performance across different hardware \cite{ant:case4codesign} and over time~\cite{karandikar:protobuf}.
Consequently, allocation decisions are often suboptimal for the neocloud as a whole as tenants start and stop instances over time.

\paragraph{Illustrative example.}
To illustrate, we evaluate two applications from different tenants: App A, a text classification fine-tuning task, and App B, a vision transformer pre-training task.
Our neocloud provides three accelerators: NVIDIA A10 and L4 GPUs and Trainium.
\autoref{tab:cost_eff} shows execution time and cost for the two applications across all three accelerators (B is incompatible with Trainium).
Assume a single A10 is available while multiple L4 and Trainium instances remain.
If A arrives first, it requests the A10 for its lowest cost (but 16\% slower), while B is left with the L4 at 8\% higher cost (and 39\% slower).
If B arrives first, it requests A10 and A is left with Trainium as its next best choice at a 13\% premium (but 14\% faster).
Viewed across both tenants, the scenario where B arrives first results in a sub-optimal allocation with higher overall cost.
Even once B terminates, A remains stuck with its suboptimal allocation.
Once allocations are made, neoclouds are unable to explore other, potentially more optimal configurations.

\begin{table}[t]
    \centering
    \begin{tabular}{cccll}
        \toprule
        \multicolumn{1}{c}{Accelerator} & \multicolumn{2}{c}{Exec. Time}  & \multicolumn{2}{c}{Exec. Cost} \\
        & A & B & \multicolumn{1}{c}{A} & \multicolumn{1}{c}{B} \\
        \midrule
        NVIDIA A10  & 0.35      & \textbf{0.23}    & \textbf{\$0.21}  & \textbf{\$0.14} \\
        NVIDIA L4   & 0.51      & 0.32    & \$0.25 (+15\%)    & \$0.15 (+8\%) \\
        Trainium    & \textbf{0.30}      & --      & \$0.24 (+13\%)    & -- \\
        \bottomrule \\
    \end{tabular}
    \caption{
        Execution time and cost (and tradeoffs) for two tenant applications across three different accelerators.}
    \label{tab:cost_eff}
\end{table}

\begin{figure}[h]%
\centering%
\includegraphics[width=0.45\textwidth]{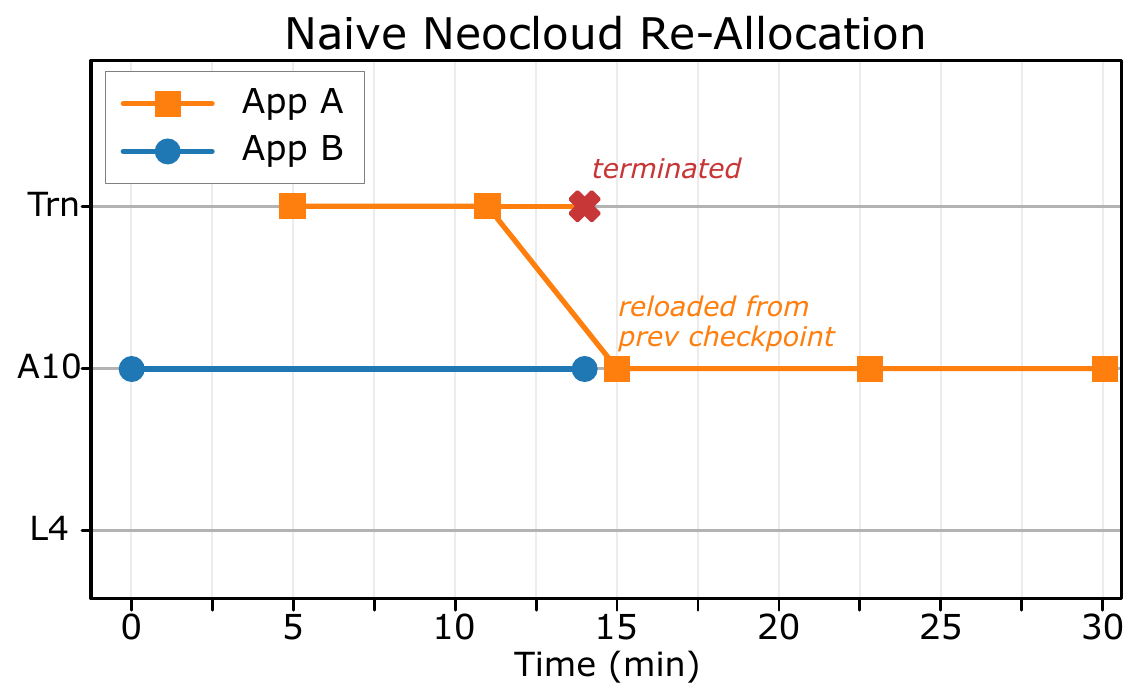}%
\caption{
    App B, is launched at minute 0 and App A, is requested at minute 5.}%
\label{fig:naive_realloc}%
\end{figure}

\paragraph{Migration improves efficiency but is challenging.}
Neoclouds lack robust migration mechanisms to leverage instance availability as it dynamically arises.
In an idealized setup with zero-overhead migration for our example above, A could migrate to the A10 instance once B terminates early.
However, migration of specialized applications across heterogeneous hardware requires highly custom implementation and incurs substantial overhead depending on the workload phase \cite{he:hype_train}.
An operator blindly migrating a tenant to a faster accelerator at the wrong time can do more harm than good.
For example, in \autoref{fig:naive_realloc} we revisit the prior scenario with a real migration implementation that incurs overhead.
Without proper coordination, the migration occurs at an inopportune moment --- towards the end of a training epoch --- resulting in substantial overhead and reduced overall efficiency.

\paragraph{Optimal allocations require dynamic coordination.}
We conclude that since the performance a tenant obtains from an accelerator fluctuates over time, the value of an instance type to a tenant — and the price they are willing to pay — also varies. 
Migrations towards optimal allocations must account for these fluctuations to arrive at ideal allocations.
However, neoclouds and clients currently lack mechanisms to negotiate pricing during execution, resulting in higher costs for clients and poor utilization for operators. 
Existing static pricing models fail to adapt to the complex real-time supply and demand dynamics of resource-constrained neoclouds.
To further complicate this, incentives of different tenants and the operators are often in conflict.

Today, operators are forced to incorporate the amortized cost of wasted capacity, which arises dynamically, into their instance pricing.
Marked-up instance prices undermine competitive advantage of neoclouds and worsen tenant's cost-per-performance.
Instead, neocloud models must efficiently align client objectives of improved cost-per-performance with operator objectives of high utilization and revenue.
The future success of neoclouds hinges on redefining the cloud-client interface to enable runtime negotiation and dynamic reallocation.

\section{Efficient Runtime Negotiation and Dynamic Allocation}%
\label{sec:insight}

While the aforementioned challenges arise from the fact that allocations are static, we note that in reality modern accelerated applications are dynamic and flexible.

Specialized applications such as DNN training are \emph{dynamic}, exhibiting phase-based hardware affinity: it may be preferable to run gradient calculations on GPUs and weight updates on FPGAs \cite{he:hype_train}.

These applications are also often \emph{flexible} and can deploy targeting a variety of hardware through high-level frameworks~\cite{abadi:tensorflow, kinzer:polymath} and just-in-time compilation \cite{software:jax2018}.

Furthermore, emerging datacenter accelerators specialize at a granularity that supports cross-domain functions (GEMM, CODECs, cryptographic hashing) \cite{santos:cerebras_mol, karandikar:cdpu, beckwith:crystals}. 
Intuitively, this flexibility in applications and hardware can be leveraged by neocloud operators for cross-accelerator migration and reallocation.
However, since leveraging this flexibility requires deep knowledge of and tight integration with application code \cite{wang:dmx, kim:aurora}, neoclouds must rely on tenants to manage their own migrations \cite{kim:dream}.
Prior work has shown how migration overheads can be greatly reduced by designing data transfer primitives into accelerators and applications \cite{suresh:mozart, ivanov:datamov}.
All that is left is for neocloud operators to appropriately align their migration objectives with incentives for tenants.

Our key insight is that tenants inherently have an understanding of their cost-per-performance tradeoff, so by renegotiating allocations through the lifetime of an application, neoclouds can incentivize tenant-driven migration that aligns the multi-party objectives of operators and other running tenants.

\textbf{Proposal:} We propose market-inspired dynamic negotiation as the core mechanism for efficient dynamic allocation in resource-constrained heterogeneous neoclouds.

Our model, \sys, enables dynamic negotiation for applications to programmatically request new allocations at runtime.
\sys employs market mechanisms to compare different applications' relative cost efficiencies, information that is visible only to individual tenants in silo.
Diversity in perceived value of hardware platforms sets this approach apart from prior market-based allocation research.
Similar to prior market-based allocation, tenants offer initial bids for hardware resources and respond to pricing changes as they run, making decisions on whether the current price for hardware is the right tradeoff between cost and performance.
However, in the \sys model, tenants and operators are responsible for characterizing their instance tradeoffs with a price.
\sys establishes pricing as the narrow waist between between operator and tenant, and among tenants to identify best-fit mappings in the unique resource-constrained and heterogeneous topology of neoclouds.
Through this channel, neoclouds can leverage price updates to incentivize utilization and fulfill their operational objectives while keeping base instance prices low.

\begin{figure}%
\centering
\includegraphics[width=0.45\textwidth]{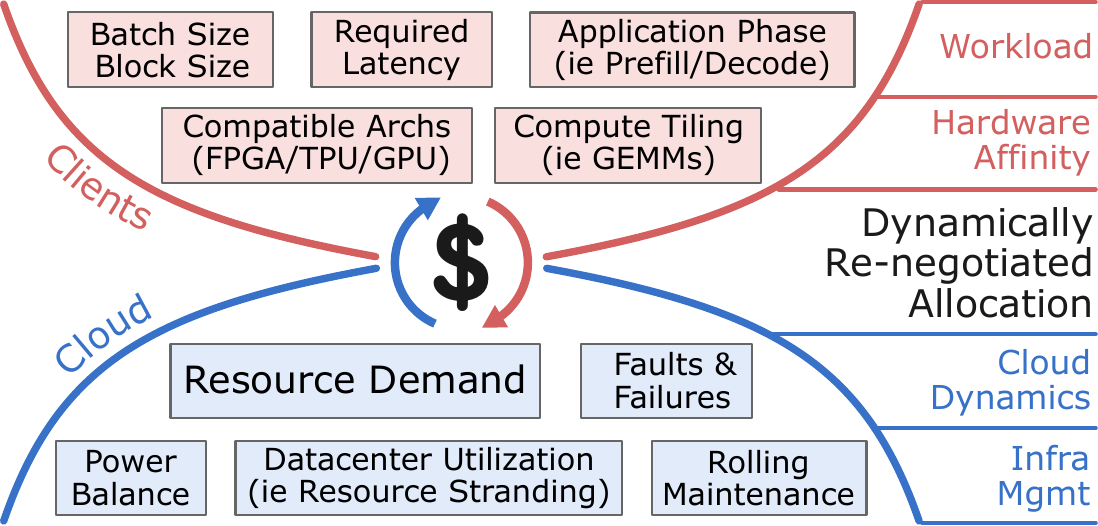}
\caption{The narrow waist between cloud operational constraints and tenant application objectives}
\label{fig:narrow_waist4}%
\end{figure}
\section{Proposed Architecture}%
\label{sec:design}

\begin{figure*}%
\centering
\includegraphics[width=\textwidth]{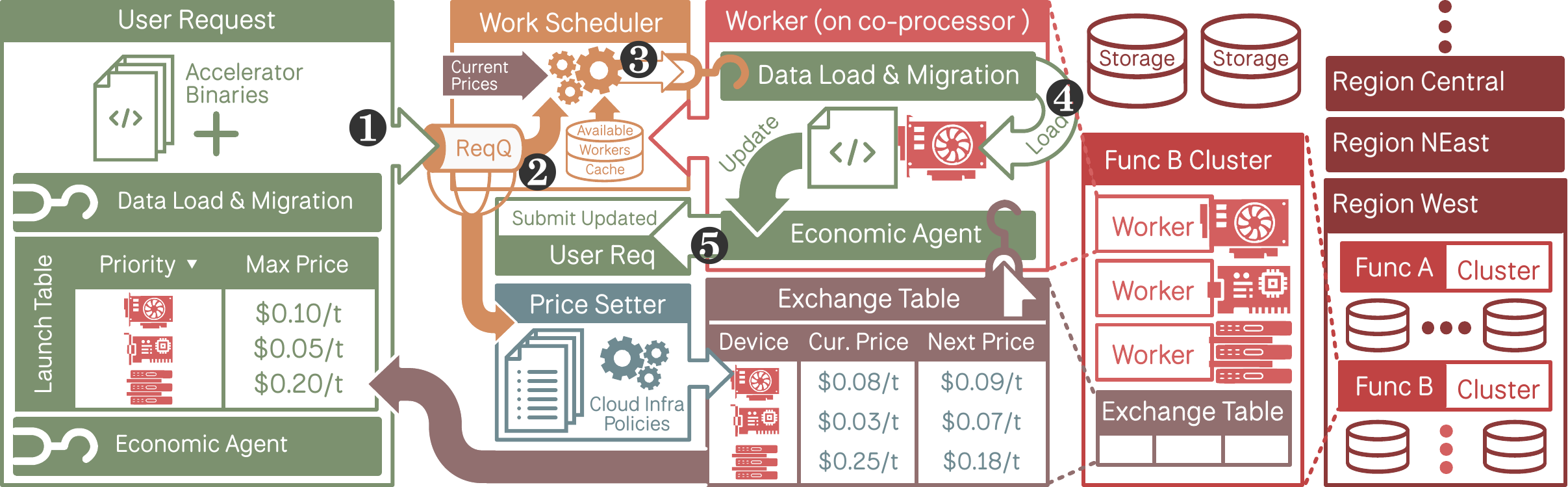}
\caption{Lifecycle of a tenant on a prototyped \sys.}%
\label{fig:lfc_arch}%
\end{figure*}

The \sys model is built on the principles of user flexibility and backwards-compatibility.

\textbf{Flexibility:} 
Tenants retain flexibility over application model and are free to submit just one, or multiple binaries for execution across various platforms.
\sys infrastructure offers hooks for tenants to host application-specific logic that can augment cloud infrastructure mechanisms for handling data movement.
Furthermore, tenants have full control over the duration of their tenancy on an instance. 
Through demand pricing, clouds are compensated for tenants who wish to occupy high-demand, limited-supply hardware.

\textbf{Backwards-compatibility:} 
Since tenants take responsibility for declaring their hardware compatibility, \sys can host new and old hardware all the same.
Furthermore, neocloud operators can confidently invest in emerging hardware as long as a substantial portion of workloads can effectively target it.
\sys{}s organize multiple accelerators, which may vary significantly in generation or vendor, into \textbf{Functional Clusters}: each dedicated to a single computational function (e.g., video encoding, cryptography, convolution, GEMMs).
Once incorporated into \sys, new hardware is harmonized with peer platforms that offer similar functionality.

Follow \autoref{fig:lfc_arch} as we outline the stages of a \sys application's lifecycle.

\circled{1} Applications begin with the submission of \textbf{User Requests} (\autoref{subsec:user_req}) to appropriate clusters, including multiple binaries for flexibility in execution across available hardware.

\circled{2} Users also submit a priority list of compatible hardwares that is evaluated against cluster supply and demand by the \textbf{Work Scheduler} (\autoref{subsec:price_n_alloc}).

\circled{3} Once a match between request and accelerator has been made, the Work Scheduler 

\circled{4} dispatches the request using tenant-provided \textbf{Data Load and Migration} hooks.

\circled{5} A tenant-provided \textbf{Economic Agent} runs in parallel, monitoring market prices of compatible cluster hardware (\autoref{subsec:exec_n_econ}) and re-queing an updated User Request to renegotiate allocation and migrate (\autoref{subsec:migration}).

Periodically, a \textbf{Price Setter} keeps the cluster-local \textbf{Exchange Table} up to date by factoring current demand and cloud operational objectives.

\subsection{User Requests}
\label{subsec:user_req}
Users requests contain:

\textbf{Execution Details}: 
A multi-binary workload or an intermediate representation (IR) describing the function to be executed, such as a TensorFlow model or kernel binary.

\textbf{Launch Table}: 
A data structure specifying 
(a) prioritized list of targetable hardware within the cluster and 
(b) maximum price the user is willing to pay for each 

\textbf{Callbacks}:  
User-provided code for managing application-specific behaviors. 
(a) Economic Agent: monitors application state and relevant pricing updates during execution 
(b) Data Load and Migration: Initializes execution on the accelerator or orchestrates state transfer when transitioning between accelerators (e.g., checkpointing or live data transfer).

\subsection{Dynamic Pricing and Allocation}
\label{subsec:price_n_alloc}
Within each functional cluster, a cluster-local exchange table maintains real-time market rates for all accelerators. 
The exchange table is updated by a price-setting agent, which factors in 
(a) Base Price: cloud-defined minimum prices for each accelerator, and 
(b) Tenant Bids: demand and outstanding bids for a given accelerator.

When a request reaches the head of the queue, the Work Scheduler evaluates it against: 
(1) Worker Availability Cache: A record of which accelerators are currently available, and 
(2) Exchange Table Prices. 
The request’s launch table is compared to current prices and if the bid for a hardware type is the highest, the Work Scheduler assigns the workload to the corresponding accelerator.

Upon successful placement, the workload is deployed to the selected accelerator for execution. 
If no match is found, the request remains in the queue, awaiting better pricing or resource availability. 
Unsatisfied requests will either time-out or can be canceled by the submitting tenant.

The Work Scheduler kicks off the application by launching the Data Load and Migration callback. 
If the application was already in execution on another platform, it migrates using mechanisms described below \autoref{subsec:migration}. 
Else, the Data Load callback performs binary loading and initialisation of the accelerator.

\subsection{Execution and Economic Behavior}
\label{subsec:exec_n_econ}
When a workload begins execution, it is accompanied by an infrastructure-layer worker process that handles events and sets up execution environment for the user’s economic agent. 
This agent operates periodically: 
(1) Monitors the cluster-local exchange table for price updates, and 
(2) Evaluates if continued execution on current accelerator is economical.

If a price increase is imminent and exceeds the user’s budget, the economic agent may 
(1) migrate the application, triggering a re-queue of the request with updated launch table for migration to a more cost-effective platform, or 
(2) terminate execution, storing/dumping intermediate state and saving cost for budget-constrained applications.

\subsection{Migration Framework}
\label{subsec:migration}
Migration callbacks handle workload-specific logic, such as: deciding when to checkpoint (e.g., at computation boundaries) or orchestrating data streaming between accelerators. 
Migration can leverage intermediate object stores or high-speed infrastructure \cite{software:gpudirect} for direct state transfer if available.

The choice between temporarily allocating source and destination accelerators or storing application state and terminating running instances is entirely up to tenants and their priorities for latency and cost. %
\section{Preliminary Results}%
\label{sec:eval}

Using our example from before, we designed simple economic agents for App A and App B where cost-to-solution increases as the workload progresses beyond the last checkpoint.
Now, when App A requests A10, it issues a bid at the break-even point of running on the next best platform.
The cost of Trainium is \$0.804/hr so the highest App A is willing to pay for A10 is \$0.687/hr (the cost of completing on A10 divided by estimated runtime on Trainium).
App B, follows a similar approach but also factors in the cost of a checkpoint restart, bidding \$0.762/hr.
App B wins by higher bid and continues running at \$0.687/hr by second-price auction rules (Vickrey–Clarke–Groves mechanism ~\cite{article:vcg}) while App A opts to progress on Trainium while keeping its bid open.
After reaching its checkpoint, App B reevaluates its bid for the A10 at \$0.652/hr factoring in estimated execution time on L4.
At this point, since A10 is still in contention for App A, App B loses the bid and migrates to L4 efficiently.
App A evaluates its state and decides to complete its epoch on Trainium, rather double-paying than losing progress.
At the next checkpoint boundary, App A migrates, loading the checkpoint onto the A10 at \$0.652/hr.
When App B eventually completes, the contention for the A10 instance is gone and the price falls to the base \$0.606/hr.

\begin{figure}%
\centering
\includegraphics[width=0.45\textwidth]{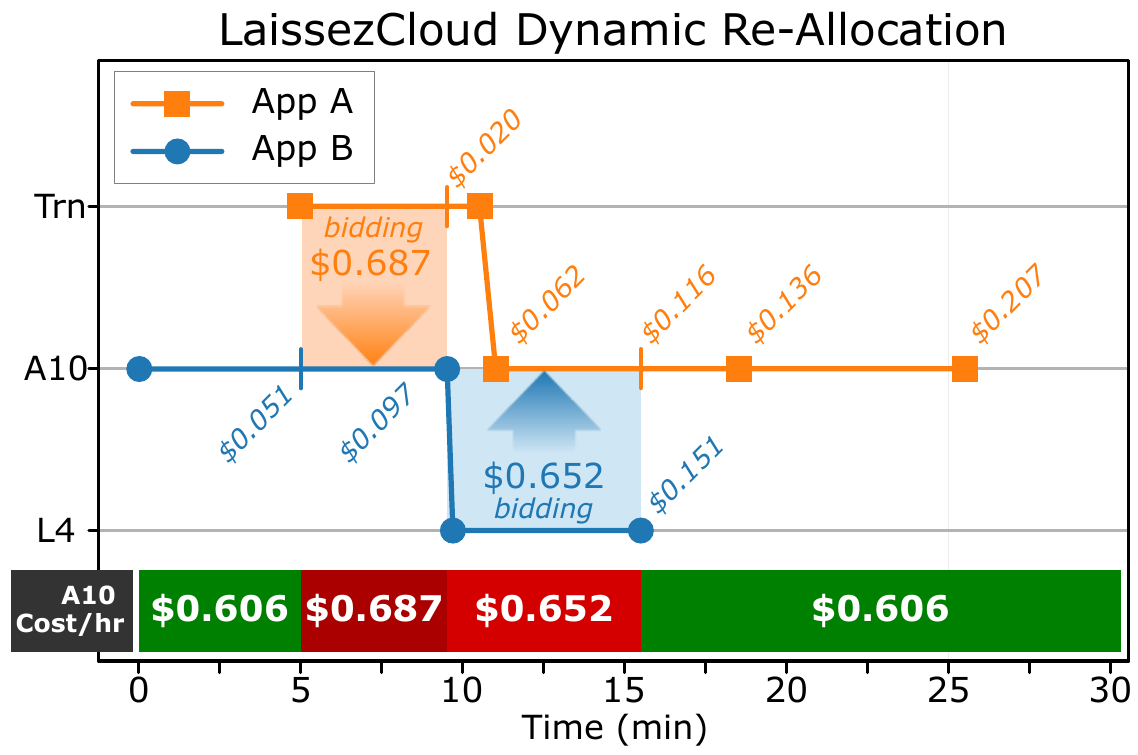}
\caption{
    App B's decision to relocate to the L4 instance is tenant-initiated based on application state and increased demand on A10.
    Annotated prices at each node represent cumulative billing.}
\label{fig:billing_laissez}%
\end{figure}

Our method of migration through checkpointing to shared storage allowed App A to pick up execution on the A10 instance within 5 seconds of allocation.
Implementing the method took only 5 lines of code, with the help of Tensorflow libraries and AWS CLI, to set up the shared storage and load/save checkpoints to it.
Results of this experiment, depicted in \autoref{fig:billing_laissez} reveal how applications can embed detailed tradeoffs into their economic models.
On the infrastructure front, free market mechanisms and leveraged application-accelerator flexibility help our neocloud achieve higher utilization.

\section{Discussion and Future Work}
\label{sec:related}
We argue that dynamic resource negotiation is essential for achieving resource and cost efficiency in neoclouds. 
With \sys, we have taken a first step toward proposing an architecture that supports this vision, providing early results to demonstrate feasibility.
However, significant challenges and opportunities remain ahead.

\paragraph{From Tenant Programming Models..}
While clouds adopt accelerator-laden disaggregated architectures, the question of how tenants should design applications remains unanswered. 
Systems like Skadi, Plaque, Ray and others \cite{hu:skadi, barham:pathways, moritz:ray, nisan:popcorn} enable execution on disaggregated accelerator clusters, but they do so at the expense of flexibility in programming models.
Forcing applications to reengineer for granular, fungible execution or specific data system integrations limits the potential market for custom hardware platforms in the cloud, posing challenges to the neocloud cost model.

Instead, we believe research should focus on enabling existing programming models to target diverse hardware. 
Such approaches \cite{mlir, leissa:anydsl, goog:xla} demonstrate extensibility and acknowledge the diversity of modern cloud workloads.

We align with prior work \cite{waldspurger:spawn, shneid:whynomark} that treats cost models as first-class citizens in cloud-deployed applications.
Integrating cost models into cloud systems bridges the gap between the economic factors that drive cloud viability and the engineering principles required for performant offerings.

\paragraph{..To Cloud Infrastructure}
While some work has prescribed new models for tenants, other work has tried bending clouds to the tenants' will \cite{zhang:userdefc}.
Compared to hyperscalers, neoclouds face greater vulnerabilities and require support from synergies across the stack.

One notable example of market-based synergy is Amazon’s 2009 spot market for CPUs.
These markets were constrained by poor incentive structures for bidding and high revocation risks caused by market volatility \cite{irwin:price}. 
Today’s accelerators, by contrast, offer an order-of-magnitude cost-per-performance advantage over general-purpose instances, leaving no viable fallback for meeting application demands.
\footnote{In the aforementioned prototype, a c8g.x16large instance of Graviton CPUs took 3 hours to complete a single epoch.}
In response, today's market-based models must convert revocation risk into revocation cost, which can be incorporated into a tenant’s strategy. 
By embedding this cost into economic models, tenants can better navigate the dynamic allocation of resources in neoclouds.

To enable such tight integrations and rapid reallocations, advancements must begin at the hardware layer. 
Work on SmartNICs ~\cite{park:lovelock} and composable streaming accelerators ~\cite{wei:cohort} shows promise in enabling rapid data migration.
Meanwhile, emerging accelerators are beginning to prioritize datacenter primitives \cite{ranga:wsvidacc, genc:gemmini} and multitenancy support \cite{kim:moca}.
Future accelerators must support stateless operations and sophisticated pipelined memory systems to facilitate efficient migration and work-stealing.
Such capabilities are foundational for realizing the full potential of neoclouds.
\section{Conclusion}
\label{sec:conclusion}

The rise of heterogeneous neoclouds with diverse accelerators—GPUs, TPUs, and FPGAs—has outpaced traditional static resource allocation models, leading to inefficiencies and poor utilization.
\sys introduces a break from the traditional IaaS model: building a bridge between cloud and client objectives and paving the way for runtime negotiation of allocations.
This paper presented a vision for the future of neoclouds: where efficient allocation is co-designed with the clients. \if \ANON 0
\section*{Acknowledgments}
We thank anonymous reviewers for their valuable feedback. 
We also thank Matheus Stolet, Vaastav Anand, and Sundar Ramamurthy for their insightful feedback and discussions.
 \fi

\bibliographystyle{plain}
\bibliography{paper,bibdb/papers,bibdb/strings,bibdb/defs}

\label{page:last}
\end{document}